\documentclass[preprint,prd,onecolumn,final,tightenlines,byrevtex]{revtex4}%
\usepackage{amsfonts}
\usepackage{amsmath}
\usepackage{amssymb}
\usepackage{graphicx}%
\begin{document}
\title{Subsolutions of the Dirac equation and WIMP dark matter}
\author{A. Okninski}
\affiliation{Physics Division, Politechnika Swietokrzyska, Al. 1000-lecia PP 7, 25-314
Kielce, Poland}

\begin{abstract}
In the present paper we study subsolutions of the Dirac equation. The known
examples of such subsolutions are the Weyl neutrino and the Majorana neutrino.
It is shown that the Dirac equation for a massive non-interacting particle
admits another decomposition into two separate equations. It is suggested that
such particles can be identified with weakly interacting massive particles,
other than neutrinos and of non-exotic nature, which can be candidates for
dark matter.\medskip



\end{abstract}
\date{\today}
\startpage{1}
\endpage{99}
\maketitle

\section{Introduction}

The Dirac equation is a relativistic quantum mechanical wave equation
formulated by Paul Dirac in 1928 providing a description of elementary
spin-$\frac{1}{2}$ particles, such as electrons and quarks, consistent with
both the principles of quantum mechanics and the theory of special relativity
\cite{Dirac1928}. The Dirac equation is \cite{Bjorken1964, Berestetskii1971,
Thaller1992}:%

\begin{equation}
\gamma^{\mu}p_{\mu}\Psi=m\Psi, \label{Dirac1}%
\end{equation}
where $m$ is the rest mass of the elementary particle, $p_{\mu}=i\frac
{\partial}{\partial x^{\mu}}$ are the components of the four-momentum
operators defined in natural units ($c=1$, $\hslash=1$). Tensor indices are
denoted with Greek letters, $\mu=0,1,2,3$ and we shall always sum over
repeated indices. The $\gamma$'s are $4\times4$ anticommuting Dirac matrices:
$\gamma^{\mu}\gamma^{\nu}+\gamma^{\nu}\gamma^{\mu}=2g^{\mu\nu}I$ where $g$
stands for the Minkowski space-time metric tensor: $g^{\mu\nu}=$%
\ \textrm{diag\thinspace}$\left(  1,-1,-1,-1\right)  $ and $I$ is a unit
matrix. The wave function is a bispinor, i.e. consists of two two-component
spinors $\xi$, $\eta$: $\Psi=\left(
\begin{array}
[c]{c}%
\xi\\
\eta
\end{array}
\right)  $.

In the present paper we study subsolutions of the Dirac equation. The known
examples of such subsolutions are the Weyl neutrino and the Majorana neutrino,
discussed in the next Section. In Section $3$ the Dirac equation for a massive
particle is decomposed into two separate equations. Lorentz covariance of
these equations is demonstrated in Section $4$ where a symmetry of
subsolutions is also found. Finally, possibility that such particles can be
identified with weakly interacting massive particles, other than neutrinos and
of non-exotic nature, is considered.

\section{Weyl and Majorana neutrinos as subsolutions of the Dirac equation}

In the $m=0$ case it is possible to obtain two independent equations for
spinors $\xi$, $\eta$\ by application of projection operators $Q_{\pm}%
=\frac{1}{2}\left(  1\pm\gamma^{5}\right)  $ to Eq.(\ref{Dirac1}) since
$\gamma^{5}\overset{df}{=}-i\gamma^{0}\gamma^{1}\gamma^{2}\gamma^{3}$
anticommutes with $\gamma^{\mu}p_{\mu}$:%
\begin{equation}
Q_{\pm}\gamma^{\mu}p_{\mu}\Psi=\gamma^{\mu}p_{\mu}\left(  Q_{\mp}\Psi\right)
=0. \label{DiracNeutrino}%
\end{equation}
In the spinor representation of the Dirac matrices \cite{Berestetskii1971} we
have $\gamma^{5}=\ \mathrm{diag\,}\left(  -1,-1,1,1\right)  $ and thus
$Q_{-}\Psi=\left(
\begin{array}
[c]{c}%
\xi\\
0
\end{array}
\right)  $, $Q_{+}\Psi=\left(
\begin{array}
[c]{c}%
0\\
\eta
\end{array}
\right)  $ and separate equations for $\xi$, $\eta$ follow:
\begin{subequations}
\label{WEYL}%
\begin{align}
\left(  p^{0}+\overrightarrow{\sigma}\cdot\overrightarrow{p}\right)  \eta &
=0,\label{Weyl1}\\
\left(  p^{0}-\overrightarrow{\sigma}\cdot\overrightarrow{p}\right)  \xi &
=0, \label{Weyl2}%
\end{align}
where $\overset{\rightarrow}{\sigma}$ denotes the vector built of the Pauli
matrices. Equations (\ref{WEYL}) are known as the Weyl equations and are used
to describe massless left-handed and right-handed neutrinos. However, since
the experimentally established phenomenon of neutrino oscillations requires
non-zero neutrino masses, theory of massive neutrinos, which can be based on
the Dirac equation, is necessary \cite{Zralek1997, Perkins2000,Fukugita2003}.
Alternatively, a modification of the Dirac or Weyl equation, called the
Majorana equation, is thought to apply to neutrinos. According to Majorana
theory neutrino and antineutrino are identical and neutral \cite{Majorana1937}%
. Although the Majorana equations can be introduced without any reference to
the Dirac equation they are subsolutions of the Dirac equation
\cite{Zralek1997}.

Indeed, demanding in (\ref{Dirac1}) that $\Psi=\mathcal{C}\Psi$ where
$\mathcal{C}$ is the charge conjugation operator, $\mathcal{C}\Psi=i\gamma
^{2}\Psi^{\ast}$, we obtain in the spinor representation $\xi=-i\sigma^{2}%
\eta^{\ast}$, $\eta=i\sigma^{2}\xi^{\ast}$and the Dirac equation
(\ref{Dirac1}) reduces to two separate Majorana equations for two-component
spinors:
\end{subequations}
\begin{subequations}
\label{MAJORANA}%
\begin{align}
\left(  p^{0}+\overrightarrow{\sigma}\cdot\overrightarrow{p}\right)  \eta &
=-im\sigma^{2}\eta^{\ast},\label{Majorana1}\\
\left(  p^{0}-\overrightarrow{\sigma}\cdot\overrightarrow{p}\right)  \xi &
=+im\sigma^{2}\xi^{\ast}. \label{Majorana2}%
\end{align}

It follows from the condition $\Psi=\mathcal{C}\Psi$ that Majorana particle
has zero charge built-in condition. The problem whether neutrinos are
described by the Dirac equation or the Majorana equations is still open
\cite{Zralek1997, Perkins2000, Fukugita2003}.

\section{Massive subsolutions of the free Dirac equation}

\subsection{The spinor representation}

The free Dirac equation (\ref{Dirac1}) in the spinor representation of
$\gamma$ matrices reads:%

\end{subequations}
\begin{equation}
\left.
\begin{array}
[c]{r}%
\left(  p^{0}+p^{3}\right)  \eta_{\dot{1}}+\left(  p^{1}-ip^{2}\right)
\eta_{\dot{2}}=m\xi^{1}\\
\left(  p^{1}+ip^{2}\right)  \eta_{\dot{1}}+\left(  p^{0}-p^{3}\right)
\eta_{\dot{2}}=m\xi^{2}\\
\left(  p^{0}-p^{3}\right)  \xi^{1}+\left(  -p^{1}+ip^{2}\right)  \xi
^{2}=m\eta_{\dot{1}}\\
\left(  -p^{1}-ip^{2}\right)  \xi^{1}+\left(  p^{0}+p^{3}\right)  \xi
^{2}=m\eta_{\dot{2}}%
\end{array}
\right\}  , \label{Dirac2}%
\end{equation}
with $\Psi=\left(  \xi^{1},\xi^{2},\eta_{\dot{1}},\eta_{\dot{2}}\right)  ^{T}$
\cite{Berestetskii1971} (see also \cite{Corson1953} for full exposition of
spinor formalism).

For $m\neq0$ we can define new quantities:
\begin{subequations}
\label{DEF1}%
\begin{align}
\left(  p^{0}+p^{3}\right)  \eta_{\dot{1}}  &  =m\xi_{(1)}^{1},\quad\left(
p^{1}-ip^{2}\right)  \eta_{\dot{2}}=m\xi_{(2)}^{1},\label{def1}\\
\left(  p^{1}+ip^{2}\right)  \eta_{\dot{1}}  &  =m\xi_{(1)}^{2},\quad\left(
p^{0}-p^{3}\right)  \eta_{\dot{2}}=m\xi_{(2)}^{2}, \label{def2}%
\end{align}

where we have:
\end{subequations}
\begin{subequations}
\label{DEF2}%
\begin{align}
\xi_{(1)}^{1}+\xi_{(2)}^{1}  &  =\xi^{1},\label{def3}\\
\xi_{(1)}^{2}+\xi_{(2)}^{2}  &  =\xi^{2}. \label{def4}%
\end{align}
In spinor notation $\xi_{(1)}^{1}=\psi_{\dot{1}}^{1\dot{1}}$, $\xi_{(2)}%
^{1}=\psi_{\dot{2}}^{1\dot{2}}$, $\xi_{(1)}^{2}=\psi_{\dot{1}}^{2\dot{1}}$,
$\xi_{(2)}^{2}=\psi_{\dot{2}}^{2\dot{2}}$.

Equations (\ref{Dirac2}) can be now written as:
\end{subequations}
\begin{equation}
\left.
\begin{array}
[c]{r}%
\left(  p^{0}+p^{3}\right)  \eta_{\dot{1}}=m\xi_{(1)}^{1}\\
\left(  p^{1}-ip^{2}\right)  \eta_{\dot{2}}=m\xi_{(2)}^{1}\\
\left(  p^{1}+ip^{2}\right)  \eta_{\dot{1}}=m\xi_{(1)}^{2}\\
\left(  p^{0}-p^{3}\right)  \eta_{\dot{2}}=m\xi_{(2)}^{2}\\
\left(  p^{0}-p^{3}\right)  \left(  \xi_{(1)}^{1}+\xi_{(2)}^{1}\right)
+\left(  -p^{1}+ip^{2}\right)  \left(  \xi_{(1)}^{2}+\xi_{(2)}^{2}\right)
=m\eta_{\dot{1}}\\
\left(  -p^{1}-ip^{2}\right)  \left(  \xi_{(1)}^{1}+\xi_{(2)}^{1}\right)
+\left(  p^{0}+p^{3}\right)  \left(  \xi_{(1)}^{2}+\xi_{(2)}^{2}\right)
=m\eta_{\dot{2}}%
\end{array}
\right\}  . \label{Dirac3}%
\end{equation}

It follows from Eqs.(\ref{DEF1}) that the following identities hold:
\begin{subequations}
\label{ID}%
\begin{align}
\left(  p^{1}+ip^{2}\right)  \xi_{(1)}^{1}  &  =\left(  p^{0}+p^{3}\right)
\xi_{(1)}^{2},\label{id1}\\
\left(  p^{0}-p^{3}\right)  \xi_{(2)}^{1}  &  =\left(  p^{1}-ip^{2}\right)
\xi_{(2)}^{2}. \label{id2}%
\end{align}

Taking into account the identities (\ref{ID}) we can finally write equations
(\ref{Dirac3}) as a system of the following two equations:%

\end{subequations}
\begin{equation}
\left.
\begin{array}
[c]{r}%
\left(  p^{0}+p^{3}\right)  \eta_{\dot{1}}=m\xi_{(1)}^{1}\\
\left(  p^{1}+ip^{2}\right)  \eta_{\dot{1}}=m\xi_{(1)}^{2}\\
\left(  p^{0}-p^{3}\right)  \xi_{(1)}^{1}+\left(  -p^{1}+ip^{2}\right)
\xi_{(1)}^{2}=m\eta_{\dot{1}}%
\end{array}
\right\}  , \label{constituent1}%
\end{equation}

\begin{equation}
\left.
\begin{array}
[c]{r}%
\left(  p^{1}-ip^{2}\right)  \eta_{\dot{2}}=m\xi_{(2)}^{1}\\
\left(  p^{0}-p^{3}\right)  \eta_{\dot{2}}=m\xi_{(2)}^{2}\\
\left(  -p^{1}-ip^{2}\right)  \xi_{(2)}^{1}+\left(  p^{0}+p^{3}\right)
\xi_{(2)}^{2}=m\eta_{\dot{2}}%
\end{array}
\right\}  . \label{constituent2}%
\end{equation}

Due to the identities (\ref{ID}) equations (\ref{constituent1}),
(\ref{constituent2}) can be cast into form:%

\begin{equation}
\left.
\begin{array}
[c]{r}%
\left(  p^{0}+p^{3}\right)  \eta_{\dot{1}}=m\xi_{(1)}^{1}\\
\left(  p^{1}+ip^{2}\right)  \eta_{\dot{1}}=m\xi_{(1)}^{2}\\
\left(  p^{0}-p^{3}\right)  \xi_{(1)}^{1}+\left(  -p^{1}+ip^{2}\right)
\xi_{(1)}^{2}=m\eta_{\dot{1}}\\
\left(  p^{0}+p^{3}\right)  \xi_{(1)}^{2}-\left(  p^{1}+ip^{2}\right)
\xi_{(1)}^{1}=0
\end{array}
\right\}  , \label{constituent1/4}%
\end{equation}

\begin{equation}
\left.
\begin{array}
[c]{r}%
\left(  p^{1}-ip^{2}\right)  \eta_{\dot{2}}=m\xi_{(2)}^{1}\\
\left(  p^{0}-p^{3}\right)  \eta_{\dot{2}}=m\xi_{(2)}^{2}\\
\left(  p^{0}-p^{3}\right)  \xi_{(2)}^{1}+\left(  -p^{1}+ip^{2}\right)
\xi_{(2)}^{2}=0\\
\left(  -p^{1}-ip^{2}\right)  \xi_{(2)}^{1}+\left(  p^{0}+p^{3}\right)
\xi_{(2)}^{2}=m\eta_{\dot{2}}%
\end{array}
\right\}  . \label{constituent2/4}%
\end{equation}

\subsection{Arbitrary representation}

Let us define the following set of mutually commuting projection operators:%

\begin{subequations}
\label{PRO}%
\begin{align}
P_{1}  &  =\tfrac{1}{4}\left(  3\mathbf{-}\gamma^{5}-\gamma^{0}\gamma
^{3}+i\gamma^{1}\gamma^{2}\right)  ,\label{P1}\\
P_{2}  &  =\tfrac{1}{4}\left(  3\mathbf{-}\gamma^{5}+\gamma^{0}\gamma
^{3}-i\gamma^{1}\gamma^{2}\right)  ,\label{P2}\\
P_{3}  &  =\tfrac{1}{4}\left(  3\mathbf{+}\gamma^{5}+\gamma^{0}\gamma
^{3}+i\gamma^{1}\gamma^{2}\right)  ,\label{P3}\\
P_{4}  &  =\tfrac{1}{4}\left(  3\mathbf{+}\gamma^{5}-\gamma^{0}\gamma
^{3}-i\gamma^{1}\gamma^{2}\right)  , \label{P4}%
\end{align}
where $\gamma^{5}=-i\gamma^{0}\gamma^{1}\gamma^{2}\gamma^{3}$,\ introduced by
Corson \cite{Corson1953} (note that in \cite{Corson1953} operators
$\epsilon_{i}=1-P_{i}$ were defined and another convention for the $\gamma$
matrices, $\gamma^{\mu}\gamma^{\nu}+\gamma^{\nu}\gamma^{\mu}=2\delta^{\mu\nu
}I$, was used). In the spinor representation of $\gamma$ matrices we have
$P_{1}=\ $\textrm{diag\thinspace}$\left(  1,1,1,0\right)  $, $P_{2}%
=\ \mathrm{diag\,}\left(  1,1,0,1\right)  $.

It is now easily checked that for $m\neq0$\ Eqs.(\ref{constituent1/4}),
(\ref{constituent2/4}) can be written in the spinor representation of the
Dirac matrices as
\end{subequations}
\begin{subequations}
\label{CONSTITUENTS4}%
\begin{align}
\gamma^{\mu}p_{\mu}P_{1}\Psi_{\left(  1\right)  }  &  =mP_{1}\Psi_{\left(
1\right)  },\label{constituent1/P}\\
\gamma^{\mu}p_{\mu}P_{2}\Psi_{\left(  2\right)  }  &  =mP_{2}\Psi_{\left(
2\right)  }, \label{constituent2/P}%
\end{align}
respectively, where%
\end{subequations}
\begin{equation}
\Psi_{\left(  1\right)  }=\left(
\begin{array}
[c]{c}%
\xi_{(1)}^{1}\\
\xi_{(1)}^{2}\\
\eta_{\dot{1}}\\
\eta_{\dot{2}}%
\end{array}
\right)  ,\quad\Psi_{\left(  2\right)  }=\left(
\begin{array}
[c]{c}%
\xi_{(2)}^{1}\\
\xi_{(2)}^{2}\\
\eta_{\dot{1}}\\
\eta_{\dot{2}}%
\end{array}
\right)  ,\quad P_{1}\Psi_{\left(  1\right)  }+P_{2}\Psi_{\left(  2\right)
}=\Psi. \label{psi}%
\end{equation}

It is important that Eqs.(\ref{constituent1/P}), (\ref{constituent2/P}) are
valid for arbitrary representation of the Dirac matrices since the projection
operators (\ref{PRO}) are defined in arbitrary representation.

Equations (\ref{constituent1}), (\ref{constituent2}) and identities (\ref{ID})
can be also given representation independent form. Indeed, from
(\ref{constituent1/P}), (\ref{constituent2/P}) the equations
\begin{subequations}
\label{CONSTITUENTS3}%
\begin{align}
P_{i}\gamma^{\mu}p_{\mu}P_{i}\Psi &  =mP_{i}\Psi,\quad\left(  i=1,2\right)
\label{constituents/3}\\
\left(  1-P_{i}\right)  \gamma^{\mu}p_{\mu}P_{i}\Psi &  =0,\quad\left(
i=1,2\right)  \label{identities}%
\end{align}
which are equivalent to Eqs.(\ref{constituent1}), (\ref{constituent2}) and
(\ref{ID}) in the spinor representation, follow, respectively. Let us note
that the identities (\ref{identities}) are the direct consequence of
Eqs.(\ref{CONSTITUENTS4}) for arbitrary representation of the Dirac matrices.

\section{Lorentz covariance and symmetries of subsolutions}

In what follows we shall use the approach of Bjorken and Drell
\cite{Bjorken1964} to study covariance of Eqs.(\ref{CONSTITUENTS4}). The
postulate of Lorentz covariance means that the two equations below are equivalent:%

\end{subequations}
\begin{equation}
\left(  i\gamma^{\mu}\frac{\partial}{\partial x^{\mu}}-m\right)  P_{k}%
\Psi\left(  x\right)  =0, \label{DP1a}%
\end{equation}

\begin{subequations}
\begin{equation}
\left(  i\tilde{\gamma}^{\mu}\frac{\partial}{\partial x^{\mu^{\prime}}%
}-m\right)  \tilde{P}_{k}^{\prime}\Psi^{\prime}\left(  x^{\prime}\right)  =0,
\label{DP2a}%
\end{equation}
where $k=1,2$. Since all representations of $\gamma\ $matrices are unitarily
equivalent \cite{Good1955} we may write Eq.(\ref{DP2a}) as:%

\begin{equation}
\left(  i\gamma^{\mu}\frac{\partial}{\partial x^{\mu^{\prime}}}-m\right)
P_{k}^{\prime}\Psi^{\prime}\left(  x^{\prime}\right)  =0. \label{DP2b}%
\end{equation}
We start with transformation of the term $P_{k}\Psi\left(  x\right)  $ in
Eq.(\ref{DP1a}):%

\end{subequations}
\begin{align}
P_{k}\Psi\left(  x\right)   &  =P_{k}S^{-1}\left(  a\right)  \Psi^{\prime
}\left(  x^{\prime}\right)  =S^{-1}\left(  a\right)  \left[  S\left(
a\right)  P_{k}S^{-1}\left(  a\right)  \right]  \Psi^{\prime}\left(
x^{\prime}\right)  =\nonumber\\
&  =S^{-1}\left(  a\right)  P_{k}^{\prime}\Psi^{\prime}\left(  x^{\prime
}\right)  , \label{PkPsi}%
\end{align}
where $S\left(  a\right)  $ is a similarity matrix depending on some set of
parameters $a$ and:%
\begin{equation}
P_{k}^{\prime}=S\left(  a\right)  P_{k}S^{-1}\left(  a\right)  . \label{P'}%
\end{equation}

Substituting (\ref{PkPsi}) into Eq.(\ref{DP1a}) we get:%

\begin{align}
S\left(  a\right)  \left(  i\gamma^{\mu}\frac{\partial}{\partial x^{\mu}%
}-m\right)  S^{-1}\left(  a\right)  P_{k}^{\prime}\Psi^{\prime}\left(
x^{\prime}\right)   &  =0,\label{DP1b}\\
S\left(  a\right)  \left(  i\gamma^{\mu}a_{\ \mu}^{\nu}\frac{\partial
}{\partial x^{\nu^{\prime}}}-m\right)  S^{-1}\left(  a\right)  P_{k}^{\prime
}\Psi^{\prime}\left(  x^{\prime}\right)   &  =0,\label{DP1c}\\
\left(  iS\left(  a\right)  \gamma^{\mu}S^{-1}\left(  a\right)  a_{\ \mu}%
^{\nu}\frac{\partial}{\partial x^{\nu^{\prime}}}-m\right)  P_{k}^{\prime}%
\Psi^{\prime}\left(  x^{\prime}\right)   &  =0. \label{DP1d}%
\end{align}
We now demand that:%

\begin{equation}
S\left(  a\right)  \gamma^{\mu}S^{-1}\left(  a\right)  a_{\ \mu}^{\nu}%
=\gamma^{\nu}. \label{Pconditions}%
\end{equation}
Equations (\ref{Pconditions}) have the following solutions:%
\begin{equation}
S=\exp\left(  -\tfrac{i}{4}\omega\sigma_{\mu\nu}I^{\mu\nu}\right)
,\qquad\sigma_{\mu\nu}\overset{df}{=}\tfrac{i}{2}\left(  \gamma_{\mu}%
\gamma_{\nu}-\gamma_{\nu}\gamma_{\mu}\right)  , \label{S}%
\end{equation}
where constant matrices $I^{\mu\nu}$ define a Lorentz transformation
\cite{Bjorken1964}.

Substituting Eq.(\ref{Pconditions}) into (\ref{DP1d}) we obtain Eq.(\ref{DP2b}%
) and this completes the proof that Eqs.(\ref{DP1a}), (\ref{DP1b}) are
equivalent. It thus follows that Eqs.(\ref{CONSTITUENTS4}) are covariant.
\medskip

There is a symmetry connecting equations (\ref{constituent1/P}),
(\ref{constituent2/P}). To show this let us notice that it is possible to
carry out Lorentz transformation to cast equations (\ref{constituent1/P}),
(\ref{constituent2/P}) into the following forms:%
\begin{equation}
\left(  \gamma^{0}p_{0}-\gamma^{1}p_{1}-m\right)  P_{1}\Psi_{\left(  1\right)
}\left(  x\right)  =0, \label{P1a}%
\end{equation}%
\begin{equation}
\left(  \gamma^{0}p_{0}-\gamma^{1}p_{1}-m\right)  P_{2}\Psi_{\left(  2\right)
}\left(  x\right)  =0, \label{P2a}%
\end{equation}
respectively, since the Lorentz transformations in the $x^{0}x^{3}$ and
$x^{1}x^{2}$ planes, are obtained from (\ref{S}) as $S_{03}=e^{-\frac{i}%
{2}\omega\sigma_{03}}$ ($I^{30}=-I^{03}=-1$), $S_{12}=e^{\frac{i}{2}%
\omega\sigma_{12}}$ ($I^{12}=-I^{21}=-1$), respectively \cite{Bjorken1964},
commute with $P_{i}$'s. Now it turns out that the unitary matrix $V$:%
\begin{equation}
V=i\gamma^{2}\gamma^{3}, \label{V}%
\end{equation}
commutes with $\gamma^{0}$, $\gamma^{1}$ and transforms $P_{1}\rightleftarrows
P_{2}$ and thus transforms Eqs. (\ref{P1a}), (\ref{P2a}) one into another in
this special frame of reference.

\section{Discussion}

It was demonstrated that, besides Weyl and Majorana neutrinos, there are also
other subsolutions of the Dirac equation, c.f. Eqs.(\ref{CONSTITUENTS4}).
These new subsolutions are covariant and are valid for arbitrary
representation of the Dirac matrices.

Derivation of Eqs. (\ref{constituent1/4}), (\ref{constituent2/4}) in the
spinor representation of the Dirac matrices was based on the identities
(\ref{ID}). Analogous identities (\ref{identities}) are fulfilled in the case
of arbitrary representation of the Dirac matrices. These identities are a
consequence of commutation of components of four-momentum operator in the free
case. It follows that in the interacting case, these equations cannot be
derived from the corresponding Dirac equation. Moreover, the mass in Eqs.
(\ref{CONSTITUENTS4}) cannot be zero and these equations are not chiral. It
thus follows that equations (\ref{CONSTITUENTS4}) describe some massive,
spin-$\frac{1}{2}$, non-interacting and non-chiral particles. It seems natural
to identify these particles with weakly interacting massive particles (WIMPs),
other than neutrinos, which are candidates for dark matter, the existence of
which was recently confirmed \cite{Massey2007}. A strong candidate for WIMP
dark matter is exotic particles, possibly the neutralino \cite{Bertone2005,
Schnee2006}. On the other hand, theory based on Eqs.(\ref{CONSTITUENTS4}) does
not invoke exotic matter.

Let us note finally that equations (\ref{constituent1/4}),
(\ref{constituent2/4}) and their $3\times3$ analogues were also found in the
Kemmer-Duffin-Petiau theory of spin $0$ and $1$ mesons \cite{Okninski2003}%
.

\end{document}